\documentclass[11pt,epsbox]{article}
\usepackage[dvips]{graphicx}
\usepackage{amsmath,amssymb}

\baselineskip=\normalbaselineskip
\renewcommand{\baselinestretch}{1.4}
\makeatletter
\@addtoreset{equation}{section}

\makeatother
\newcommand{\be}{\begin{equation}}
\newcommand{\ee}{\end{equation}}
\newcommand{\ba}{\begin{eqnarray}}
\newcommand{\ea}{\end{eqnarray}}
\newcommand{\nn}{\nonumber \\}

\newcommand{\del}{\partial}
\newcommand{\bra}[1]{\left\langle\,{#1}\,\right|}
\newcommand{\ket}[1]{\left|\,{#1}\,\right\rangle}
\newcommand{\bracket}[2]{
\left\langle\left.\,{#1}\,\right|\,{#2}\,\right\rangle}

\newcommand{\Tr}{{\rm Tr}}

\begin{document}
\setcounter{page}{0}
%%%%%%%%%%%%%%%%%%%

\begin{flushright}
\parbox{40mm}{%
TU-855 \\
%RIKEN-TH-47 \\
%hep-th/yymmnnn \\
November 2009}
\end{flushright}

\vfill

%%%%%%%%%%%%%%%%%%%%%%%%%%%%%%%%%%%%%%%%%%%%%%%%%%%%
%% Title
%%%%%%%%%%%%%%%%%%%%%%%%%%%%%%%%%%%%%%%%%%%%%%%%%%%%
\begin{center}
{\Large{\bf 
Noncommutative Solitons of Gravity \\
% or \\
% Emergence of Spacetime via Noncommutativity (tentative)
}}
\end{center}

\vfill

\renewcommand{\baselinestretch}{1.0}

%%%%%%%%%%%%%%%%%%%%%%%%%%%%%%
%% author
%%%%%%%%%%%%%%%%%%%%%%%%%%%%%%
\begin{center}
\textsc{Tsuguhiko Asakawa}
\footnote{E-mail: \texttt{asakawa@tuhep.phys.tohoku.ac.jp}} 
and  
\textsc{Shinpei Kobayashi}
\footnote{E-mail: \texttt{shimpei@nat.gunma-ct.ac.jp}} 
%\\[2em]

~\\
$^{1}$ \textsl{Department of Physics, Graduate School of Science, \\
Tohoku University, \\
Sendai 980-8578, JAPAN} \\ 

\vspace{0.5cm}
$^2$ \textsl{Department of Physics, \\
      Gunma National College of Technology,  \\
     580 Toribacho, Maebashi, 371-8530, JAPAN}  
\end{center}

%%%%%%%%%%%%%%%%%%%%
%% abstract
%%%%%%%%%%%%%%%%%%%%
\begin{center}
{\bf abstract}
\end{center}

\begin{quote}

\small{%
We investigate a three-dimensional gravitational theory 
on a noncommutative space which has 
a cosmological constant term only. 
We found various kinds of nontrivial solutions
by applying a similar technique which was used to seek 
noncommutative solitons in noncommutative scalar field theories. 
Some of those solutions correspond to bubbles of spacetimes 
or represent dimensional reduction. 
The solution which interpolates $G_{\mu\nu}=0$ and the Minkowski metric 
is also found. 
All solutions we obtained are non-perturbative 
in the noncommutative parameter 
$\theta$, 
therefore they are different from solutions found in other 
contexts of noncommutative theory of gravity 
and would  have a close relation to quantum gravity. }
\end{quote}
\vfill

\renewcommand{\baselinestretch}{1.4}

%%%%%%%%%%%%%%%%%%%%%%%%%%%%%%%%%%
% Main
%%%%%%%%%%%%%%%%%%%%%%%%%%%%%%%%%%
\renewcommand{\thefootnote}{\arabic{footnote}}
\setcounter{footnote}{0}
\addtocounter{page}{1}
%%%%%%%%%%%%%%%%%%%%%%%%%%%%%%%%%
\newpage
\section{Introduction}
\label{sec:intro}

The construction of a consistent theory of spacetime at the Planck 
scale is one of the main issue in fundamental physics. 
There is an expectation that a noncommutativity among spacetime coordinates, 
\be
[x^\mu, x^\nu] = i\theta^{\mu\nu},
\ee
emerges in such a scale. 
In fact, there are so many attempts at making this idea manifest, 
that is, to construct a consistent theory of gravity with a noncommutativity 
to be taken into account. 
For example, a noncommutative extension of the gauge theory of gravitation 
has been investigated \cite{Chamseddine:2000si, Moffat:2000gr}. 
This formalism is based on gauging the noncommutative $SO(1,4)$ de Sitter 
group \cite{Zet:2003bv} and using the Seiberg-Witten map \cite{Seiberg:1999vs}
with subsequent contraction to the Poincar\'{e} group $ISO(1,3)$. 
In that theory,  corrections to cosmological and black hole solutions 
due to the noncommutativity have been found \cite{Mukherjee:2007fa, 
Chaichian:2007we, Chaichian:2007dr}. 
Application of the Seiberg-Witten map to Chern-Simons 
theoires have been carried out in 
\cite{Mukherjee:2004cb, Mukherjee:2006pf}.
Utilizing the correspondence between three-dimensional Einstein gravity 
and three-dimensional Chern-Simons theory, %shares a similar idea on which 
the noncommutative gauge theory of gravitation is considered in 
\cite{Witten:1988hc, Banados:2001xw, ChangYoung:2008my, Kim:2007nx}. 
Another approach to noncommutative spacetimes is 
considering noncommutative effects on gravitational sources 
\cite{Nicolini:2009gw, Nicolini:2008aj, Nicolini:2005zi, Nicolini:2005vd, Ansoldi:2006vg, Spallucci:2008ez, 
Myung:2008kp, Banerjee:2008du, Banerjee:2008gc, Banerjee:2009gr, Banerjee:2009xx}. 
The authors have found some solutions which solve the Einstein equation 
with gravitational sources of Gaussian type whose widths are related to the noncommutativity. 
This approach is directly connected to an expectation of smearing curvature 
singularities that appear in Einstein gravity. 
Also, the authors of \cite{Aschieri:2005zs, Aschieri:2005yw}
proposed a theory of gravity on noncommutative spaces 
from the viewpoint of twisting the diffeomorphism.  
This theory has been extended to a theory which includes 
fermionic terms, i.e., a supergravity on noncommutative spaces 
\cite{Aschieri:2009ky, Aschieri:2009mc}. 
There are some trials to give classical solutions for those theories
and actually a few solutions have been found 
\cite{Aschieri:2009qh, Schupp:2009pt, Schenkel:2009qm, Ohl:2009pv}.
Other approaches to noncommutative gravity also can be found 
in \cite{Mukherjee:2006nd, Banerjee:2007th, Vassilevich:2009cb}.

Although these approaches are different in the basic hypothesis, 
they are aiming to construct a consistent noncommutative 
gravitational theory by deforming the Einstein-Hilbert action 
by the noncommutative parameter $\theta$, that comes back to the 
ordinary Einstein gravity in the commutative limit $\theta\to0$.
Moreover, the solutions already found are also deformations of solutions 
for the ordinary Einstein  gravity, 
namely, we have not had any nontrivial solutions 
particular to the gravitational theories on noncommutative spaces so far. 

In this paper we take a rather different approach to investigate the effect 
of a noncommutativity, by finding classical solutions 
that can not be obtained by deformations of solutions 
of commutative theories but are non-perturbative in the noncommutativity $\theta$.
To this end, we would like to work with a three-dimensional 
noncommutative gravitational theory that consists of a cosmological constant term only, 
that is to say, a noncommutative gravitational theory without the Ricci scalar. 
We adopt the first-order (vielbein) formalism,  
and the action of our theory reduces to 
just the three-dimensional determinant of the vielbein.
One reason to work with this situation is that the cosmological constant term is 
made by the $\star$-multiplication only and that would be common for many approaches 
to noncommutative gravity, i.e., it is model independent. 

This set-up is also motivated by the idea of \cite{Gopakumar:2000zd},
where some noncommutative solitons have been derived 
in noncommutative scalar field theories. 
The authors of \cite{Gopakumar:2000zd} take a limit that 
the space noncommutativity is very large, which makes
the kinetic term negligible compared with the potential term 
of that scalar field theory. 
Since all derivatives disappear, 
we naively expect that we can not find nontrivial solutions, 
but this is not the case due to the noncommutativity. 
Actually they found some classically stable solutions called 
noncommutative solitons. 
Soon after \cite{Gopakumar:2000zd}, their theory was extended 
to that which includes the kinetic term, and by a solution-generating 
technique, the solutions which solve the equation
of motion including the kinetic term have been explicitly 
constructed \cite{Harvey:2000jb}. 
Our case, as the determinant is made of the $\star$-multiplication of vielbein, 
is analogous to the noncommutative $\phi^3$-theory  
investigated in \cite{Gopakumar:2000zd}, and we can apply a similar technique 
to find nontrivial solutions, namely, by switching to 
the operator formulation and using projection operators or their generalization.
One of the purposes of this paper is to construct such noncommutative solitons 
of gravity.

By comparing to the noncommutative scalar solitons,
our theory is regarded naturally as the situation where the scalar curvature can be 
negligible in comparison with the cosmological constant,
but we will argue that there is another possibility in which 
the theory would be interpreted in a more radical way as the emergence of 
spacetime only from the cosmological constant term without the scalar curvature.
In any case, the solutions we found suggest a close connection to quantum gravity,
where degenerate metrics play important roles.
Such a degenerate metric that satisfies $\det E_\mu^a =0$ or $\det G_{\mu\nu}=0$ 
represents a non-classical phase of the theory and contributes to the path integral.
In particular, the diffeomorphism invariant phase $E_\mu^a =0$ is considered as 
unbroken vacuum, while the metricity condition does not restrict the 
spin connection $\omega^{ab}_\mu$ to be the Christoffel symbol, 
which becomes a completely independent variable.
This implies that the first-order formalism using vielbein is not 
equivalent to the ordinary second-order formalism using metrics.
Another characteristic feature of quantum gravity is that  
topology and signature-changing solutions are allowed 
\cite{Horowitz:1990qb}. 
The solutions obtained in this paper share the same features as above.

The organization of this paper is as follows: in the following section, 
we give our action and derive the equation of motion.
In section 3, we will construct solutions for the equation 
of motion by using projection operators. 
We first give examples to show typical structures of the solutions
(bubbles of spacetime, dimensional reduction).
Then the most general solutions of this class are presented.
In section 4, we construct another class of solutions by using 
Gamma matrices, which are more close to conventional spacetimes.
The final section is devoted to discussion and future directions.

%------------------%
\section{The Noncommutative Gravity of Cosmological Constant}

\subsection{Action and Equation of Motion}
Let us start with a three-dimensional noncommutative plane 
${\mathbb R}^3$ with coordinates $x^\mu (\mu=0,1,2)$ or $(t,x,y)$.  
The star product is defined for any functions on ${\mathbb R}^3$ as
\begin{equation}
\left(f \star g \right)(x) = \exp\left(\frac{i}{2}
\frac{\del}{\del x^{\mu}} \theta^{\mu\nu} \frac{\del}{\del y^{\nu}}\right)
f(x)g(y)\big|_{y\to x}\ ,
\label{star-product}
\end{equation}
where $\theta^{\mu\nu}$ is a constant, anti-symmetric matrix 
which represents a noncommutativity. 
In this paper, for simplicity,
%in order to avoid a conceptual problem of causality, 
we introduce the noncommutativity purely 
in the spatial coordinates\footnote{
In general, we can choose one of coordinates which remains commutative by 
changing $\theta^{\mu\nu}$ to the Jordan form.  
Time direction is usually chosen in order to avoid infinite time derivatives. 
However, for {\it static} classical solutions, 
(\ref{star-product}) reduces automatically to (\ref{NCdef}) only. 
}  
\begin{equation}
[x, y]_{\star} 
\equiv x\star y -y\star x 
= i\theta,
\label{NCdef}
\ee
by choosing $\theta^{0i}=0 \,(i=1,2)$ and $\theta^{12}\equiv\theta$.

We exploit the first-order formulation of a three-dimensional theory of gravity 
on a noncommutative ${\mathbb R}^3$ 
which has a cosmological constant term only, 
\be
\label{theory}
S = -\frac{\Lambda }{\kappa^2}
\int dt d^2x  \ E^{\star},
\ee
where $\Lambda$ is a cosmological constant. 
Here $E^{\star}$ is the $\star$-determinant defined by 
\be
\label{detE}
E^{\star} = \mbox{det}_{\star} E
= \frac{1}{3!}\epsilon^{\mu\nu\rho} \epsilon_{abc} 
E_{\mu}^a  \star E_{\nu}^b  \star E_{\rho}^c,  
\ee
where $E_{\mu}^a (x)$ is a vielbein. 
We denote spacetime indices by $\mu, \nu, \rho$ and 
tangent space indices by $a, b, c$. 
All indices run from $0$ to $2$.  
The metric is also defined through the star product 
in a similar way \cite{Banados:2001xw, Aschieri:2005zs}:
\be 
\label{G}
G_{\mu\nu} = \frac{1}{2}\left(E_{\mu}^a\star E_{\nu}^b 
+ E_{\nu}^b \star E_{\mu}^a\right) \eta_{ab},
\ee
where $\eta_{ab}$ is an $SO(1,2)$ invariant metric of the local Lorentz frame.
We do not assume that $E_{\mu}^a$ or $G_{\mu\nu}$ are invertible as 
$3\times 3$ matrices, that is, we allow degenerate metrics.
Through this paper, $G_{\mu\nu}$ is assumed to be real for simplicity. 
The solutions we discuss later will not contradict with this assumption, 
but complex metrics can also be treated in a similar manner.

Here we would like to point out that 
there are two possibilities (a) and (b) to see this simple setting 
in a full gravitational theory on the noncommutative space:
(a) The action given in (\ref{theory}) is a part of a full theory, 
that is, we need to add 
a noncommutative generalization of the Einstein-Hilbert term
to (\ref{theory}).
This is of course the common belief.
In this case, our theory (\ref{theory}) is considered to be valid 
when the scalar curvature term is 
negligible compared with the cosmological constant.
However, as opposed to the noncommutative scalar field theory, 
this is not achieved by taking the large noncommutativity limit $\theta\to \infty$ 
of a certain full theory
\footnote{ 
We recall the argument in \cite{Gopakumar:2000zd}:
by rescaling the coordinates $x\to x/\sqrt{\theta},\ 
y\to y/\sqrt{\theta}$, all $\theta$ in the star product disappear,
while any other derivatives (and also gauge fields) acquire $1/\sqrt{\theta}$.
Then all the derivative terms become negligible in the large $\theta$ limit.
However, as opposed to the scalar theory, the vielbein and other quantities 
in (\ref{EH theory})
are also transformed under the rescaling keeping the action invariant.
Thus, the simple rescaling argument cannot be directly applied 
to our case.
Note also that we should take a static or a slowly time-varying approximation
as well, in order to drop time derivatives.
}
\be
S = \frac{1}{2\kappa^2}
\int dt d^2x  \ E^{\star}\star 
\left( R_{\star} - 2\Lambda \right),
\label{EH theory}
\ee
where $R_{\star}$ is a suitably defined scalar curvature, which may be model dependent.

On the other hand, we propose another possibility in this paper:
(b) The action given in (\ref{theory}) is already a full theory.
In this case, the metric and other quantities like a scalar curvature 
are considered to be composite quantities made from the vielbein.
For a quantity in ordinary Einstein theory 
we can define several different quantities in the noncommutative case. 
For example, another metric rather than (2.5) can be defined by 
$g_{\mu\nu}=E^a_\mu \cdot E^b_\nu \eta_{ab}$ using ordinary product\footnote{
This is possible because a product $f\cdot g$ of two functions is also written by the
$\star$-product \cite{Aschieri:2005zs}.
To this end, first apply the bi-differential exponential operator 
inverse to that appears in (\ref{star-product}) to $f$ and $g$, 
then take the $\star$-product.
}.
We call the latter a ``commutative" metric in this papaer, but do not confuse! 
It is just a quantity in the noncommutative theory. 
Both ``commutative" and ``noncommutative" quantities are used for 
capturing the spacetime structures given by a classical solution of the vielbein.
In this paper, we will use two kinds of determinants 
${\rm det} G$ and ${\rm det}_\star G$ of the metric (\ref{G}), 
and ``commutative" scalar curvatures.
In this way, we switch effectively from the first-order (vielbein) formalism
to the second-order (metric) formalism without introducing a spin connection.
We emphasize that the noncommutativity makes it possible.
Such a kind of prescription would have never appeared 
in the literature to our knowledge.
This is motivated by the disagreement between the first and the second-order
formalism in phases with degenerate metrics in quantum gravity.
Of course this possibility itself should be justified, but we will see 
in this paper that the solutions 
in this interpretation possess interesting properties, suggesting a
connection to the very notion of quantum gravity.

Now let us derive the equation of motion of our theory (\ref{theory}).
We use the fact that the cyclic permutation of the star product 
is allowed in the integral:
\ba
\label{cyclic}
\int f\star g\star h &=& \int f(g\star h) =\int (g\star h)f \nonumber \\
&=& \int g\star h\star f,  
\ea
which comes from a property of the star product 
\be
\int f\star g = \int g\star f =\int fg.   
\ee
Taking this into account, 
the action (\ref{theory}) is rewritten as    
\ba
S 
&=& - \lambda \int dt d^2x \ \epsilon_{abc} E_{0}^a  
\{ E_{1}^b , E_{2}^c\}_\star, 
\label{action}
\ea
where $\lambda =\frac{\Lambda }{\kappa^2}$.
%up to the propotional coefficient. 
Here we used 
the star-anti-commutator defined by
$
\{f,g\}_{\star} \equiv f\star g+g\star f.
$
Varying the action (\ref{action}) with respect to $E_\mu^a$ 
and using the cyclic symmetry of the star product, 
we have nine equations of motion 
for ${}^\forall \mu$ and ${}^\forall a$,
\be
\epsilon^{\mu\nu\rho}\epsilon_{abc}\{E_{\nu}^b, E_{\rho}^c\}_{\star}=0.
\label{eom}
\ee

Clearly the action (\ref{action}) will be zero if the vielbein 
solves (\ref{eom}), that is, all classical solutions give  
degenerate vielbein that satisfies $\det_\star E=0$.
Nevertheless, as we will explicitly show, there are in fact 
nontrivial solutions other than $\det G=0$. %$E^a_\mu=0$.
This is contrast to the theory only with the cosmological constant term
defined on a commutative space, where only  
$\det G=0$ is allowed due to the absence of the kinetic term.  
This is because the star product has an infinite number of 
derivatives in it, which act as an effective kinetic term.

\subsection{Star Product and Operator Formulation}
In the following sections we explicitly give solutions of Eq.(\ref{eom}).
For simplicity, we will consider static or stationary solutions there.
In order to find solutions, we exploit the recipe used in \cite{Gopakumar:2000zd}, i.e., 
the usage of the connection between the star product and the operator formulation,
an analogue of the Weyl-Wigner correspondence in quantum mechanics 
(see also \cite{Harvey:2001yn} for a review). 
The vielbein $E^a_\mu(x,y)$ is a function on ${\mathbb R}^2$ if it is static.
Recall that, given a (suitably defined) function $f(x,y)$ on ${\mathbb R}^2$, 
there is a map which uniquely assigns to it an operator 
$O_{f}(\hat{x},\hat{y})$ that acts on the corresponding one-dimensional quantum 
mechanical Hilbert space $\mathcal{H}=L^2({\mathbb R})$ 
with $[\hat{x}, \hat{y}]=i\theta$. 
By choosing the Weyl ordering prescription, the Weyl map is given by 
\be
O_f(\hat{x},\hat{y}) = \frac{1}{(2\pi)^2}
\int d^2k \tilde{f}(k) \ e^{i(k_x \hat{x} + k_y \hat{y})},
\ee
where
\be
\tilde{f}(k)=\int d^2x \ e^{-i(k_x x + k_y y)}f(x,y)
\ee
is the Fourier transformation.
Then the algebra of functions with the $\star$-multiplication is isomorphic to the 
operator algebra with relations
\begin{align}
&O_f \cdot O_g = O_{f\star g}. \\
&\Tr \,O_f = \int \frac{d^2x}{2\pi \theta} f.
\end{align}
The creation and the annihilation operator are defined by 
\be
\hat{a} = \frac{\hat{x}+i\hat{y}}{\sqrt{2\theta}}, \quad
\hat{a}^{\dagger} = \frac{\hat{x}-i\hat{y}}{\sqrt{2\theta}}.
\label{HO}
\ee
The Hilbert space $\mathcal{H}$ is now spanned by orthonormal basis
$\ket{n}\, (n=0,1,2,\cdots)$, which is the energy eigenstate of the one-dimensional 
harmonic oscillator given in (\ref{HO}). 
Thus a general operator $O$ acting on $\mathcal{H}$ can be written as 
the linear combination of the matrix elements of the form
\be
O=\sum_{i,j=0}^{\infty} O^i_j \ket{i}\bra{j}.
\ee
In particular, the projection operator $\ket{i}\bra{i}$ 
will be important to construct solutions in the following sections. 
The function (symbol) $\phi_i$ corresponding to 
the projection operator (that is $O_{\phi_i}=\ket{i}\bra{i}$) 
can be expressed as \cite{Gopakumar:2000zd, Harvey:2001yn}
% \be
% \phi_i (x,y)
% = \frac{1}{2\pi}\int d^2k e^{-k^2/4}L_i\left(\frac{k^2}{2}\right) e^{-ik\cdot x}
% =2(-1)^i e^{-r^2}L_i\left(2r^2\right),
% \label{laguerre}
% \ee
\be
\phi_i (x,y)
=2(-1)^i e^{-r^2/\theta}L_i\left(\frac{2r^2}{\theta}\right),
\label{laguerre}
\ee
where $L_i(x)$ is the $i$th Laguerre polynomial 
and $r^2 = x^2 + y^2$.
By construction, $\phi_i$ is the orthogonal projection
\be
\phi_i \star \phi_j = \delta_{ij} \phi_i
\label{projection}
\ee
and satisfy the completeness relation\footnote{
By using the generating function for the Laguerre polynomials
\be
\sum_{i=0}^{\infty} L_i^{(\alpha)}(x)t^i
=\frac{1}{(1-t)^{\alpha +1}}\exp \left(-\frac{xt}{1-t}\right),
\ee
it is shown explicitly: 
\be
\sum_{i=0}^{\infty}\phi_i 
=2e^{-r^2/\theta}
\sum_{i=0}^{\infty} (-1)^i L_{i} \left(\frac{2r^2}{\theta}\right) =1.
\ee
}
\be
\sum_{i=0}^{\infty} \phi_i =1.
\label{completeness}
\ee
In the following, we sometimes use a loose notation not to distinguish $O_f$ 
and $f$.

%---------------------%
\section{Noncommutative Solitons by Projection Operators}

In this section, we will give various 
static solutions of the equation of motion (\ref{eom})
 using projection operators.
We begin by simple solutions of two types and then 
move to more general solutions.

%%%%%%%%%%%%%%%%%%%%%%%%%%%%
\subsection{Diagonal Solution}

As a warm-up, let us first consider the case with a diagonal vielbein, namely, 
we take an ansatz 
\ba
E_{\mu}^{a}
&=&
\left(
\begin{array}{ccc}
E_0^0  & 0  & 0 \\
0 & E^1_1  & 0  \\
0 & 0  & E_2^2
\end{array}
\right)
\label{diagonal E}
\ea
as a $3\times 3$ matrix.
In this case, the equation of motion (\ref{eom}) reduces to three equations 
\ba
E_0^0:&&0=\{ E_{1}^1 , E_{2}^2\}_\star, \nonumber\\
E_1^1:&&0=\{ E_{2}^2 , E_{0}^0\}_\star, \nonumber\\
E_2^2:&&0=\{ E_{0}^0 , E_{1}^1\}_\star. 
\label{diagonal eom}
\ea
Therefore, if each component of the vielbein is given by 
a projection operator and they are orthogonal among them, 
then they solves (\ref{diagonal eom}).
The simplest choice is
\ba
\label{simplest solution}
E_{\nu}^{b}
&=&
\left(
\begin{array}{ccc}
\alpha_0 \phi_0 & 0 & 0  \\
0 & \alpha_1 \phi_1 & 0 \\
0 & 0 & \alpha_2 \phi_2
\end{array}
\right), 
\ea
where $\alpha_0, \alpha_1$ and $\alpha_2$ are 
arbitrary complex numbers.
Of course, any other choice of three different projection operators 
(say $\phi_3$, $\phi_{16}$, and $\phi_{51}$, etc.) is also a solution.
More generally, arbitrary mutually orthogonal three groups of  
projection operators are allowed.

This simple example already possesses some interesting features, as we will see below.
In order to give insight to the solution (\ref{simplest solution}), 
we apply the prescription announced in the previous section to this example.
First, we can see that all solutions of this type give non-zero metric $G_{\mu\nu}$.
In fact, (\ref{simplest solution}) gives the following line element:
\ba
ds^2 
&=& G_{\mu\nu} dx^{\mu}dx^{\nu} \nonumber\\
&=& \frac{1}{2}\left(E_{\mu}^a\star E_{\nu}^b 
+ E_{\nu}^b \star E_{\mu}^a\right) \eta_{ab} dx^{\mu}dx^{\nu} \nonumber\\
&=& -\alpha_0^2 \phi_0^{2\star} dt^2 
+ \alpha_1^2 \phi_1^{2\star} dx^2 
+ \alpha_2^2 \phi_2^{2\star} dy^2 \nonumber\\
&=& -\alpha_0^2 \phi_0 dt^2 
+\alpha_1^2 \phi_1 dx^2 
+\alpha_2^2 \phi_2 dy^2 \nonumber\\
&=& 2e^{-r^2/\theta} \Big(-\alpha_0^2 L_0 (2r^2/\theta)dt^2 
-\alpha_1^2 L_1(2r^2/\theta)dx^2 
+ \alpha_2^2 L_2(2r^2/\theta)dy^2 \Big)\nonumber\\
&=& 2e^{-r^2/\theta} 
\left(- \alpha_0^2 dt^2 -\alpha_1^2 \left(1-\frac{2r^2}{\theta}\right)dx^2 
+\alpha_2^2 \left(1-\frac{4r^2}{\theta} + \frac{2r^4}{\theta^2} \right)dy^2 \right),
\label{diagonal_metric}
\ea
where we used the property of the projection operators (\ref{projection}). 
We clearly see that the metric becomes singular when we take the commutative limit 
$\theta \to 0$. 
This means that this solution can not exist if we start from the commutative 
theory. 
Furthermore, for finite $\theta$, 
other ``commutative" quantities defined from this metric (\ref{diagonal_metric}) 
are now computable,
because it is non-degenerate in the sense that it gives $\det G \neq 0$ except for some points.
Of course, as mentioned in the previous section, this treatment itself 
should be justified.

By adopting the remark above, we can evaluate the 
Ricci scalar $R$ and  the Kretschmann invariant 
$R_{\mu\nu\rho\sigma}R^{\mu\nu\rho\sigma}$ from (\ref{diagonal_metric}) 
by the standard analysis. 
The results are shown in figures \ref{fig1} and \ref{fig2}. 
The explicit forms of them are given in Appendix \ref{Ricci_and_Kretschmann}.
\begin{figure}[t]
  \begin{center}
    \includegraphics[scale=0.65]{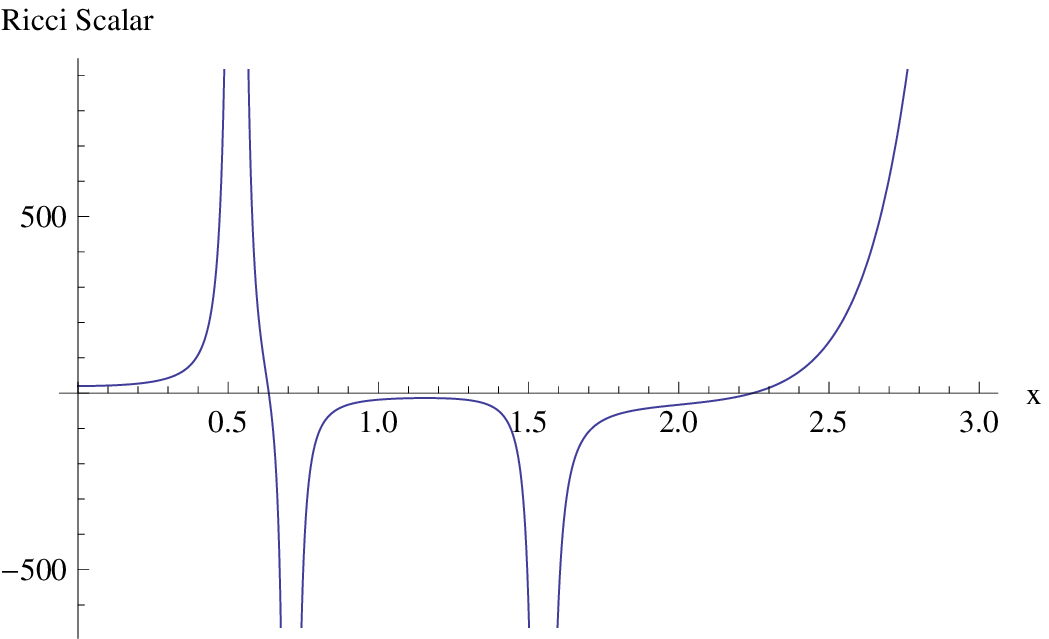}
    \quad
    \includegraphics[scale=0.65]{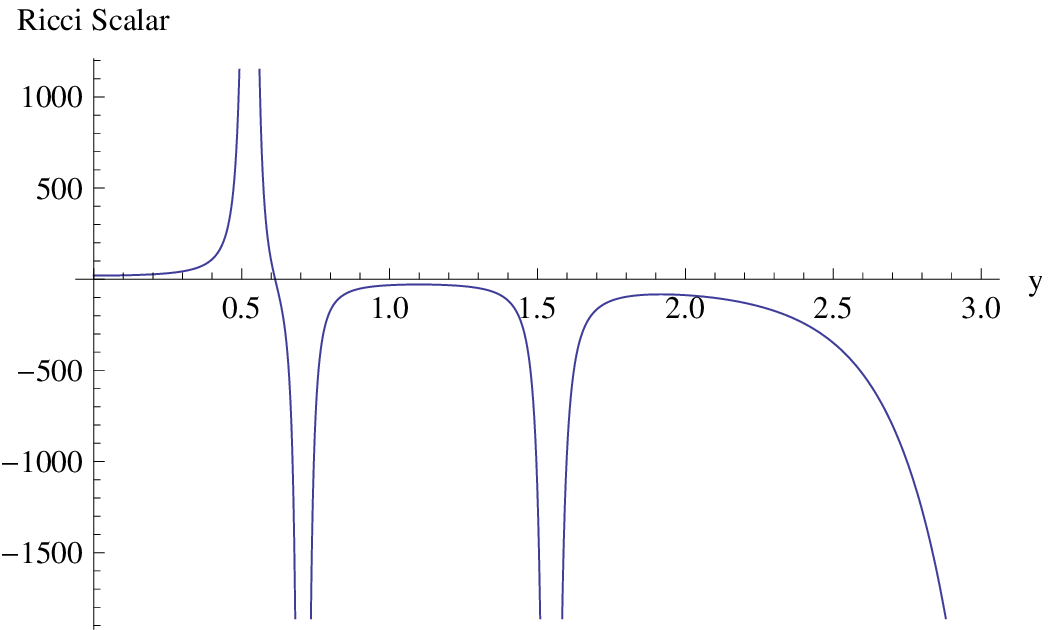}
  \end{center}
  \caption{The value of the Ricci scalar $R$ of spacetime (\ref{diagonal_metric}).  
  The left and right graphs are the $y=0$ and the $x=0$ section of $R$, respectively. 
  Here we set $\theta=1$ and $\alpha_0=\alpha_2=1/\sqrt{2}$ and $\alpha_1 =i/\sqrt{2}$ 
  as an example. } 
  \label{fig1}
\end{figure} 
\begin{figure}[t]
  \begin{center}
    \includegraphics[scale=0.65]{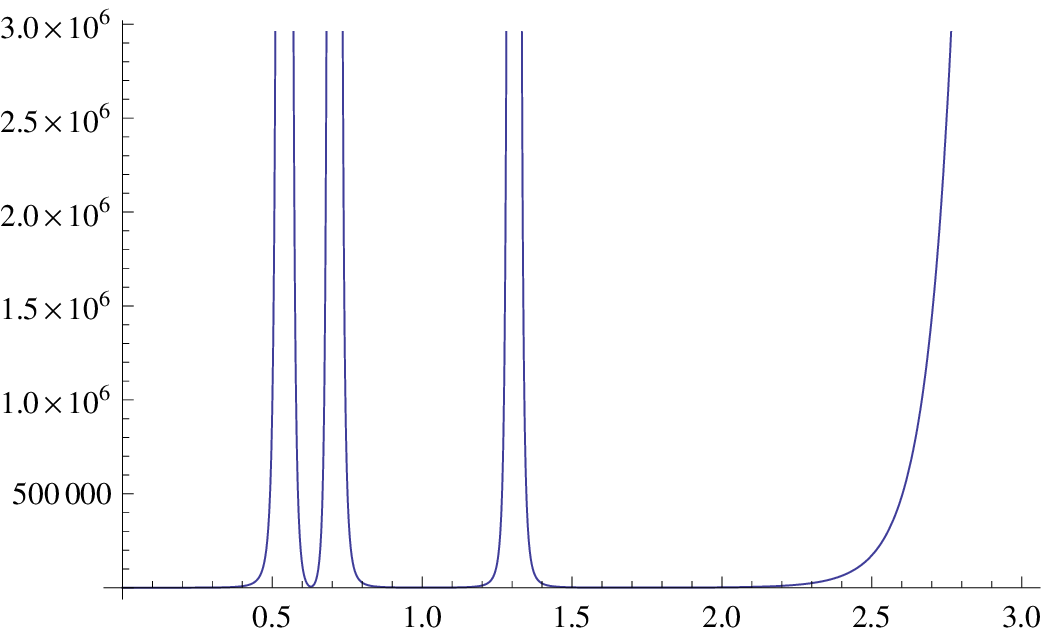}
    \quad
    \includegraphics[scale=0.65]{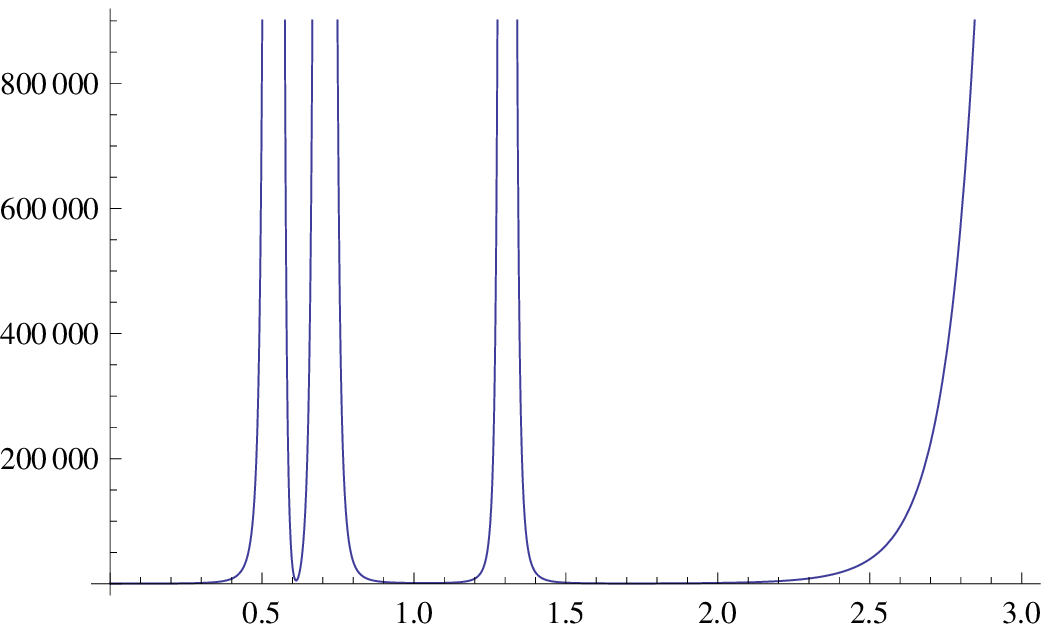}
  \end{center}
  \caption{The value of the Kretschmann invariant 
  $R_{\mu\nu\rho\sigma}R^{\mu\nu\rho\sigma}$ of spacetime (\ref{diagonal_metric}).  
  The left and right graphs are the $y=0$ and the $x=0$ section of 
  $R_{\mu\nu\rho\sigma}R^{\mu\nu\rho\sigma}$, respectively. 
  We set $\theta=1$ and $\alpha_0=\alpha_2=1/\sqrt{2}$ and $\alpha_1 =i/\sqrt{2}$ 
  as well as Figure \ref{fig1}. } 
  \label{fig2}
\end{figure} 
All of them diverge at $r=\infty$ coming from the overall factor  $e^{-r^2/\theta}$ 
appeared in 
(\ref{diagonal_metric}), and also diverge at several values of $r$ which comes from 
the zero points of the Laguerre polynomials.
As seen from these figures, the spacetime is divided into several radial regions by
the walls of curvature singularities.
The divergent points agree with those satisfy $\det G=0$.
In each region, the Ricci scalar evaluated by ordinary GR method is 
meaningful because $\det G\ne 0$.
The result is not exactly but very closed to $0$, and moreover, is almost constant 
for finite $\theta$. 
Because $\theta $ is a free parameter, we can take a commutative limit $\theta \to 0$.
Then we see that all of the walls shrink to $r=0$, 
and the space measured by the metric concentrates to one point with a curvature singularity.
Conversely, the metric at the finite $\theta$ can be viewed as a resolution 
of such a ``one-point spece".
This solution suggests that the bubbles of several spacetimes with small cosmological constants 
would emerge as a fine structure of a single point.
This fact might give a new direction for the cosmological constant problem.
\footnote{
Note also the signature of the metric.
Due to the nature of the Laguerre polynomials, the sign of each component of the metric 
oscillates as $r$ increases. 
This is not surprising because 
such a sign-changing solution is also typical in the black hole spacetime,
where in the interior of the event horizon $dt^2$ becomes spacelike while 
$dr^2$ becomes timelike.
The signs of the coefficients of $dt^2 $ and $dr^2$ 
changes independently in our solution.}

\subsection{Nondiagonal Solutions and Dimensional Reduction}

Next, let us slightly generalize the above and take a non-diagonal ansatz
for the vielbein of the form 
\ba
&&\left(
\begin{array}{ccc}
E_0^0 & 0 &0 \\
0 & E_1^1 & E_1^2 \\
0 & E_2^1 & E_2^2
\end{array}
\right).
\ea
The equation of motion (\ref{eom}) reduces to five equations
\ba
&&0=\{ E_{1}^1 , E_{2}^2\}_\star - \{ E_{1}^2 , E_{2}^1\}_\star, \\
&&0=\{ E_{0}^0, E_{\mu}^a\}_\star \quad (a, \mu = 1, 2). 
\ea
We will give solutions that represent effectively two-dimensional spacetime.

For example, we can easily find a solution which consists of the two projections 
$\phi_0$ and  $\phi_1$ as
\ba
E_{\nu}^{b}
&=&
\left(
\begin{array}{ccc}
\alpha_0\phi_0 & 0 & 0  \\
0 & \alpha_1\phi_1 & \alpha_1\phi_1 \\
0 & \alpha_1\phi_1 & \alpha_1\phi_1
\end{array}
\right),
\label{dim red 1}
\ea
where $\alpha_0$ and $\alpha_1$ are arbitrary constants as before.
This implies the metric\footnote{
We assume $dxdy=dydx$.}
\ba
ds^2 
&=& -\alpha_0^2\phi_0 dt^2 + 2\alpha_1^2\phi_1 \big( dx^2 +2dxdy + dy^2\big) \\
&=& 2e^{-r^2/\theta} \Bigg( -\alpha_0^2 dt^2 
-2\alpha_1^2 \left(1-\frac{2r^2}{\theta}\right) ( dx + dy)^2 \Bigg).
\ea
As seen in the second term of the metric, the line element 
effectively consists of $dt$ and $dx+dy$.
In other words, the metrical dimension of this metric is $2$. 
The disagreement between the naive (manifold) dimension and the metrical dimension 
would  be a sign of quantum gravity again \cite{Horowitz:1990qb}. 
In particular, it would be interesting to compare it with 
the results obtained in the analysis by causal dynamical triangulation
\cite{Ambjorn:2004qm, Ambjorn:2005db, Ambjorn:2005qt, Ambjorn:2005jj} 
or spontaneous dimensional reduction 
in short-distance quantum gravity 
\cite{Carlip:2009kf}.

A similar solution only with a single projection operator $\phi_0$ is
obtained from the above solution by replacing
$\phi_0 \to 1-\phi_0$ and $\phi_1 \to \phi_0$.
Its metric is given by
\ba
ds^2 
&=& -(1-\phi_0) dt^2 + 2\phi_0 (dx^2 +2dxdy + dy^2) \\
&=& -\left(1-2e^{-r^2/\theta} \right)dt^2 + 4e^{-r^2/\theta}  ( dx + dy)^2.
\label{dim red 2}
\ea
This is again an effectively two-dimensional metric.

On the other hand, effectively one-dimensional solutions are
obtained in the most general ansatz and the corresponding equation of motion (\ref{eom}).
For example, by using a single projection operator $\phi_0$, the vielbein
\ba
E_{\nu}^{b}
&=&
\left(
\begin{array}{ccc}
\phi_0 & \phi_0 & \phi_0  \\
\phi_0 & \phi_0 & \phi_0 \\
\phi_0 & \phi_0 & \phi_0
\end{array}
\right)
\label{dim red 3}
\ea
solves Eq.(\ref{eom}).
The line element of this solution
\ba
ds^2 
&=& \phi_0 (dt+dx+dy)^2 \\
&=& 2e^{-r^2/\theta} (dt+dx+dy)^2.
\ea
shows that the metric effectively reduces 
to a one-dimensional metric.  
The disagreement between the naive dimension and the metrical dimension 
appears again. 
Clearly this happens because of the degeneracy of the vielbein. 
In other words, the rank or the invertibility of the vielbein 
determines whether such a dimensional 
reduction occurs or not. 
We discuss this point again in the following subsection.

\subsection{General Solutions by Projection Operators}
The structure of the dimensional reduction in above examples 
suggest a systematic construction of solutions.
In general, each component of vielbein is a function on noncommutative ${\mathbb R}^2$ 
(we refer time-independent metrics only) 
and is written as an operator acting on the Hilbert space of 
a harmonic oscillator.
Therefore, the most general expression of the vielbein is written as
\be
E_{\mu}^a = \sum_{i,j=0}^{\infty} (C_{\mu}^{a})^i_j \ket{i}\bra{j},
\ee
where $(C_{\mu}^{a})^i_j$ is a (complex) number.
Now a (star) product of two components is written by using $\bracket{j}{k}=\delta^{j}_{k}$
as a matrix multiplication for $i,j$:
\be
E_{\mu}^a \star E_{\nu}^b= \sum_{i,j=0}^{\infty} (C_{\mu}^{a}C_{\nu}^{b})^i_j \ket{i}\bra{j}.
\ee
Thus, the metric is given by using the anti-commutator as
\be
G_{\mu\nu}= \frac{1}{2} 
\eta_{ab} \sum_{i,j=0}^{\infty} \{C_{\mu}^{a},C_{\nu}^{b}\}^i_j \ket{i}\bra{j}.
\ee
Similarly, the determinant (for $\mu$ and $a$) 
\begin{align}
\det (E_{\mu}^a) &= \sum_{i,j=0}^{\infty} \left[\det (C_{\mu}^{a})\right]^i_j 
\ket{i}\bra{j} 
% &= \frac{1}{3!} \epsilon_{abc}  \sum_{i,j=0}^{\infty} 
% \left[C_{0}^{a}\{C_{1}^{b}, C_{2}^{c}\}\right]^i_j 
% \ket{i}\bra{j}
\end{align}
reduces to the determinant of the matrix $C_{\mu}^{a}$.
Correspondingly, the equation of motion (\ref{eom}) reduces to the following 
constraint for $(C_{\mu}^a)_i^j$
\be
\epsilon^{\mu\nu\rho}
\epsilon_{abc} (C_{\nu}^b C_{\rho}^c)_j^i =0. 
\label{constraint2}
\ee

As a particular situation, 
let us assume the diagonality in $i,j$, that is, 
each vielbein is written in the linear 
combination of the projection operators as 
\be
E_{\nu}^b = \sum_{j=0}^{\infty} C(j)_{\nu}^{b} \phi_j, 
\quad C(j)_{\nu}^{b}\equiv (C_{\nu}^{b})^j_j ~.
\label{linear combi}
\ee
Then, (\ref{constraint2}) becomes
\be
\epsilon^{\mu\nu\rho}
\epsilon_{abc} 
C(j)_{\nu}^{b} C(j)_{\rho}^{c} =0,
\label{cofactor}
\ee
for an arbitrary $j$ (no summation).
For a fixed $j$, this is an ordinary (commutative) 
matrix equation and $C(j)_{\nu}^{b}$ 
is seen as a $3\times 3$ matrix for $\nu$ and $b$
\footnote{
It is also equivalent to the equation of motion for the vielbein $e_{\nu}^{b}$
in the commutative theory 
that has only one matrix $C$.},
\ba
C(j)
&=&
\left(
\begin{array}{ccc}
C(j)_{0}^{0}  & C(j)_{1}^{0}  &C(j)_{2}^{0}  \\
C(j)_{0}^{1}  & C(j)_{1}^{1}  &C(j)_{2}^{1}  \\
C(j)_{0}^{2}  & C(j)_{1}^{2}  &C(j)_{2}^{2}
\end{array}
\right).
\ea
Then this constraint simply shows that all minors (the determinants of 
cofactor matrices) of each matrix element $C(j)_{\nu}^{b}$ 
should be zero.
The most general form of such a matrix is given by 
\ba
C(j)
&=&
\left(
\begin{array}{c}
\alpha_j \\
\beta_j  \\
\gamma_j
\end{array}
\right) 
\left(
\begin{array}{ccc}
s_j & t_j & u_j
\end{array}
\right) 
=
\left(
\begin{array}{ccc}
\alpha_j  s_j & \alpha_j t_j & \alpha_j u_j  \\
\beta_j  s_j & \beta_j  t_j & \beta_j u_j  \\
\gamma_j s_j & \gamma_j t_j & \gamma_j u_j  
\end{array}
\right)
\label{sol C(j)}
\ea
where $\alpha_j, \beta_j, \gamma_j, s_j, t_j$ and $u_j$ are arbitrary constants. 
This means that $C(j)$ is a matrix whose rank is 1, parametrized by ${\mathbb C}^6$.
Here, the remarkable fact is that any linear combination (\ref{linear combi}),
with each $C(j)$ given by (\ref{sol C(j)}),
is also a solution due to the orthogonality of $\phi_j$'s.
Therefore, we can in fact generate an infinite number of classical solutions 
easily by assigning a set of degenerate matrices $\{C(j)\}_{j \in {\mathbb Z}}$.
We conclude that the most general solution of the vielbein and  
the corresponding metric written by the projection operators are as follows:
\ba
E_{\mu}^a
&=& 
\left(
\begin{array}{ccc}
E_0^0 & E_0^1 & E_0^2 \\
E_1^0 & E_1^1 & E_1^2 \\
E_2^0 & E_2^1 & E_2^2 
\end{array}
\right)
=
\sum_{j=0}^\infty
\left(
\begin{array}{ccc}
\alpha_j  s_j & \alpha_j t_j & \alpha_j u_j  \\
\beta_j  s_j & \beta_j  t_j & \beta_j u_j  \\
\gamma_j s_j & \gamma_j t_j & \gamma_j u_j  
\end{array}
\right) \phi_j,  
\label{general vielbein}\\
G_{\mu\nu}
&=&
\left(
\begin{array}{ccc}
G_{00} & G_{01} & G_{02} \\
G_{10} & G_{11} & G_{12} \\
G_{20} & G_{21} & G_{22} 
\end{array}
\right)
=
\sum_{j=0}^\infty
\left(-\alpha_j^2 + \beta_j^2 +\gamma_j^2\right) 
\left(
\begin{array}{ccc}
s_j^2 & s_j t_j & s_j u_j  \\
s_j t_j & t_j^2 & t_j u_j  \\
s_j u_j  & t_j u_j & u_j^2  
\end{array}
\right) \phi_j.
\label{general metric}
\ea
It is immediately shown that any metric of the form 
(\ref{general metric}) satisfies $\det_\star G=0$.

Of course, all the solutions obtained so far are characterized in this way.
In fact, the first example giving (\ref{diagonal_metric}) is characterized by
the following three degenerate matrices
\be
C(0)
=
\left(
\begin{array}{ccc}
\alpha_0 & 0 & 0  \\
0 & 0 & 0 \\
0 & 0 & 0
\end{array}
\right), \quad
C(1)
=
\left(
\begin{array}{ccc}
0 & 0 & 0  \\
0 & \alpha_1 & 0 \\
0 & 0 & 0
\end{array}
\right), \quad
C(2)
=
\left(
\begin{array}{ccc}
0 & 0 & 0  \\
0 & 0 & 0 \\
0 & 0 & \alpha_2, 
\end{array}
\right).
\label{3d example}
\ee 
As we mentioned before, 
the fact that some solutions have the discrepancy between the dimension of 
the manifolds and that of the metrics can be explained by 
the degeneracy of these matrices.
The examples (\ref{dim red 1}) and (\ref{dim red 3}) given in the previous subsection 
are characterized by
\be
C(0)
=
\left(
\begin{array}{ccc}
\alpha_0 & 0 & 0  \\
0 & 0 & 0 \\
0 & 0 & 0
\end{array}
\right), \quad
C(1)
=
\left(
\begin{array}{ccc}
0 & 0 & 0  \\
0 & \alpha_1 & \alpha_1 \\
0 & \alpha_1 & \alpha_1
\end{array}
\right),
\ee
and 
\be
C(0)
=
\left(
\begin{array}{ccc}
\alpha_0 & \alpha_0 & \alpha_0  \\
\alpha_0 & \alpha_0 & \alpha_0 \\
\alpha_0 & \alpha_0 & \alpha_0
\end{array}
\right), 
\ee
respectively.
Since each matrix $C(j)$ carries rank $1$, the sum of two such terms 
in the former gives effective two dimension, while the latter gives one dimension.
In other words, we need at least three non-zero matrices $C(j)$ in order to 
construct a three-dimensional solution as (\ref{3d example}).
As noted above, the commutative theory corresponds to a single matrix $C$.
Along the argument here, it is clear that 
the metric in the commutative theory is at most one dimensional.

In summary, even for restricting the diagonal (projection) operators in $i,j$,
we have found infinitely many solutions characterized by the infinite set 
of degenerate matrices $C(j)$.
Dividing by the symmetry, we would obtain the vacuum moduli space of the theory
in this diagonal sector.
%{\bf The structure is $\{O(3)/SO(3)\}^{\mathbb Z}$?}

We close this section by a remark.
Although we consider the three-dimensional theory in this paper, 
the extension to the $(2n+1)$-dimensional theory is straightforward.
Then the construction of the solutions in this section is also applied to the higher 
dimensional case.
To be more precise, the vielbein is represented as operators on the $n$-dimensional 
harmonic oscillator basis $\ket{j_1,j_2,\cdots,j_n}$.
The eom $\epsilon^{\mu_1,\dots,\mu_n}\epsilon_{a_1,\cdots,a_n} 
E^{a_1}_{\mu_1}\star\cdots \star E^{a_n}_{\mu_n}=0$ is solved in the same way as 
(\ref{general vielbein}) but now the sum is over any projection operators 
$\phi_{j_1,\cdots,j_n}=\ket{j_1,j_2,\cdots,j_n}\bra{j_1,j_2,\cdots,j_n}$,
because each matrix $C(j_1,j_2,\cdots,j_n)$ is independently solved similarly to
(\ref{sol C(j)}).
The structure of the dimensional reduction is also same,
which means that the invertibility of the vielbein  might be a key to
the mechanism of compactification of higher-dimensional theories.

%%%%%%%%%%%%%%%%%%%%%%%%%%%%%%%%%%%%
%%%%%%%%%%%%%%%%%%%%%%%%%%%%%%%%%%%%
\section{Noncommutative Solitons by Clifford Algebras}

In this section, we will give another class of solutions represented by
various dimensional Clifford algebras.
Here all solutions are proportional to the Minkowski metric and satisfy 
$\det_\star G\ne 0$, as opposed to the solutions in \S 3.

\subsection{First Solution}
Let us come back to the ansatz (\ref{diagonal E}) for the vielbein. 
In the previous section, we found solutions using the projection operators, 
which correspond to the diagonal matrix elements $\ket{i}\bra{i}$ 
in the harmonic oscillator basis.
However, because the equation of motion (\ref{diagonal eom}) shows that 
the vielbein should be mutually anti-commuting, 
the vielbein obeying the Clifford algebra relation solves (\ref{diagonal eom}).
Such a solution is generally represented by a non-diagonal matrix 
element $\ket{i}\bra{j}$ in that basis.

To be more precise, let us for example focus on the indices $i=0,1$ 
and define the $SO(3)$ gamma matrices (Pauli matrices) as
\begin{align}
&\gamma^0 = \sigma^3 = \ket{0}\bra{0}-\ket{1}\bra{1}, \nonumber\\
&\gamma^1 = \sigma^1 = \ket{1}\bra{0}+\ket{0}\bra{1}, \nonumber\\
&\gamma^2 = \sigma^2 = i\ket{1}\bra{0}-i\ket{0}\bra{1}.
\end{align}
They satisfy the Clifford algebra relation 
$\{\gamma^\mu, \gamma^\nu\}=2\delta^{\mu\nu} {\bf 1}_2$.
Here, ${\bf 1}_2=\ket{0}\bra{0}+\ket{1}\bra{1}$ is a unit matrix
in the two-dimensional subspace spanned by $\ket{0}$ and $\ket{1}$,
which is equivalent to the projection operator $\phi_0 +\phi_1$ 
in the full Hilbert space.
Then the vielbein of the form 
\ba
E_{\mu}^{a}
&=&
\left(
\begin{array}{ccc}
\gamma^0  & 0  & 0 \\
0 & \gamma^1  & 0  \\
0 & 0  & \gamma^2
\end{array}
\right)
\ea
is evidently a solution for (\ref{diagonal eom}).
The metric for this vielbein is 
\begin{align}
G_{\mu\nu}&= \eta_{\mu\nu}  \left(\ket{0}\bra{0}+\ket{1}\bra{1}\right)\nonumber\\
&=\eta_{\mu\nu}  \left(\phi_0 + \phi_1  \right) \\
&= \frac{4r^2}{\theta} e^{-r^2/\theta} \eta_{\mu\nu}. 
\label{metric_gamma}
\end{align}
In the last line, we rewrote $\phi_0$ and $\phi_1$ 
in terms of the Laguerre polynomials.  

Remarkably, this metric is proportional to the three-dimensional Minkowski metric,
so that it is natural to regard this solution as a soliton
that interpolates two vacua $G_{\mu\nu}=0$ and $G_{\mu\nu}=\eta _{\mu\nu}$.
The overall factor of the projection operators means that the (noncommutative) 
Minkowski space exists only in the region where $\phi_0+\phi_1$ has non-zero 
support
in analogy to the interpretation of the noncommutative scalar solitons:
On the noncommutative plane, each projection $\phi_i$ 
shares a region with a minimal area $2\pi \theta$, which is 
determined by the uncertainly relation.
It is indeed seen by noting 
$\det_\star G = (\det \eta) \left(\phi_0+\phi_1 \right)_\star^3 
= -\left(\phi_0+\phi_1 \right)$ 
and 
$\Tr \left(\phi_0+\phi_1 \right)=2$.
This implies that an effective cosmological constant term defined by ${\rm det}_\star G$ 
(it is a composite quantity different from our action) is given by 
\be
S_{eff}=-\lambda \int\! dtd^2 x \sqrt{-\mbox{det}_\star G }
%= -2\lambda \int\! dtd^2x \left(\phi_0+\phi_1 \right)
= -2\pi \theta \lambda \int\! dt \Tr \left(\phi_0+\phi_1 \right)
= -4\pi \theta \lambda \int\! dt ,
\label{volume of metric_gamma}
\ee
which means the finite volume $4\pi \theta$ in the spatial direction.
Because now $\det_\star G\ne 0$, it is in principle possible to compute the 
noncommutative scalar curvature $R_\star$ to capture the structure further.
But it needs a proper definition of $R_\star$ of course, and we will not perform it 
in this paper.

Nevertheless, the same qualitative feature can be observed 
in analyzing ``commutative" quantities.
In this treatment, the support of $\phi_0 +\phi_1$ (non-degenerate region) as a function 
is $0<r<\infty$. 
However, note that $r$ is the radial coordinate in the isotropic coordinates 
\be
ds^2 = \frac{4r^2}{\theta} e^{-r^2/\theta} (-dt^2 + dr^2 + r^2 d\varphi^2),
\label{matrixgrav1}
\ee
which differs from the physical radial coordinate 
usually defined by $R = (2r^2/\sqrt{\theta}) e^{-r^2/2\theta}$.
Both $r=0$ and $r=\infty$ correspond to $R=0$.
% (see Fig.\ref{fig:radius_diagonal_metric}). 
This implies that the physical distance between $r=0$ and $r=\infty$ is  finite.
It is consistent with the finite spatial integral in (\ref{volume of metric_gamma}).
With this in mind, we now check invariant scalars
$R$
and $R_{\mu\nu\rho\sigma}R^{\mu\nu\rho\sigma}$ 
for this metric, 
and the results are given as (see also Fig.\ref{fig:rs_diagonal_metric})
\ba
&& R = -\frac{e^{r^2/\theta}}{2r^4 \theta}(\theta^2-6r^2\theta +r^4), \\
%&& R_{\mu\nu}R^{\mu\nu}
%= \frac{e^{2r^2}}{8r^8}(3-11r^2+28r^4-9r^6+r^8)\\
&& R_{\mu\nu\rho\sigma}R^{\mu\nu\rho\sigma}
= \frac{e^{2r^2}}{4r^8\theta^2}(5\theta^4
-10r^2\theta^3+18r^4\theta^2-6r^6\theta+r^8).
\ea
\begin{figure}[t]
  \begin{center}
    \includegraphics[scale=0.65]{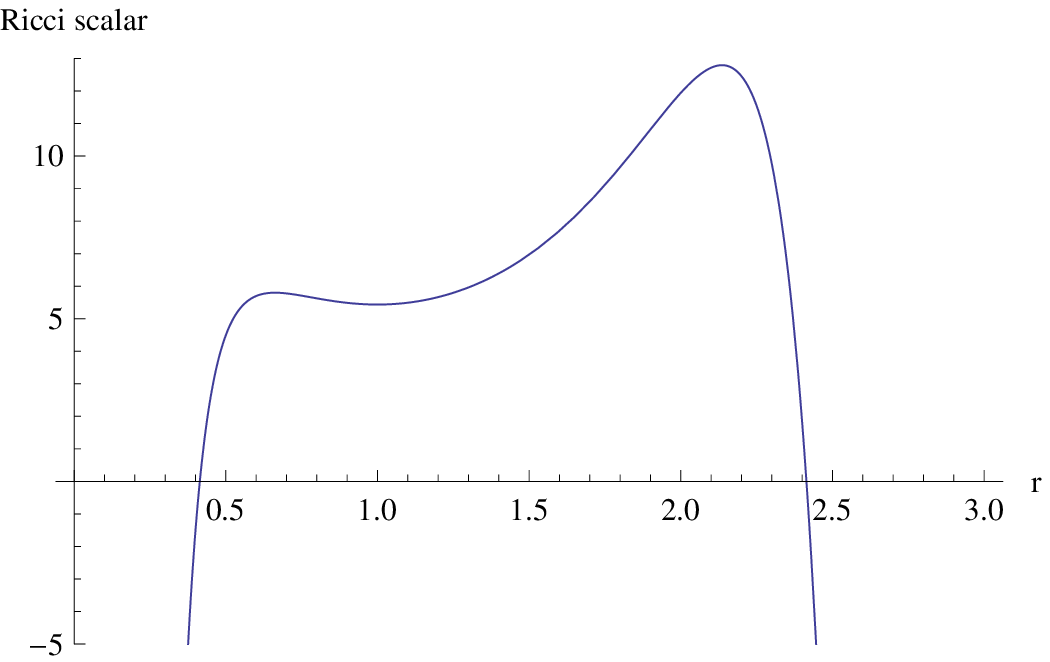}
    \hspace{1cm}
    \includegraphics[scale=0.65]{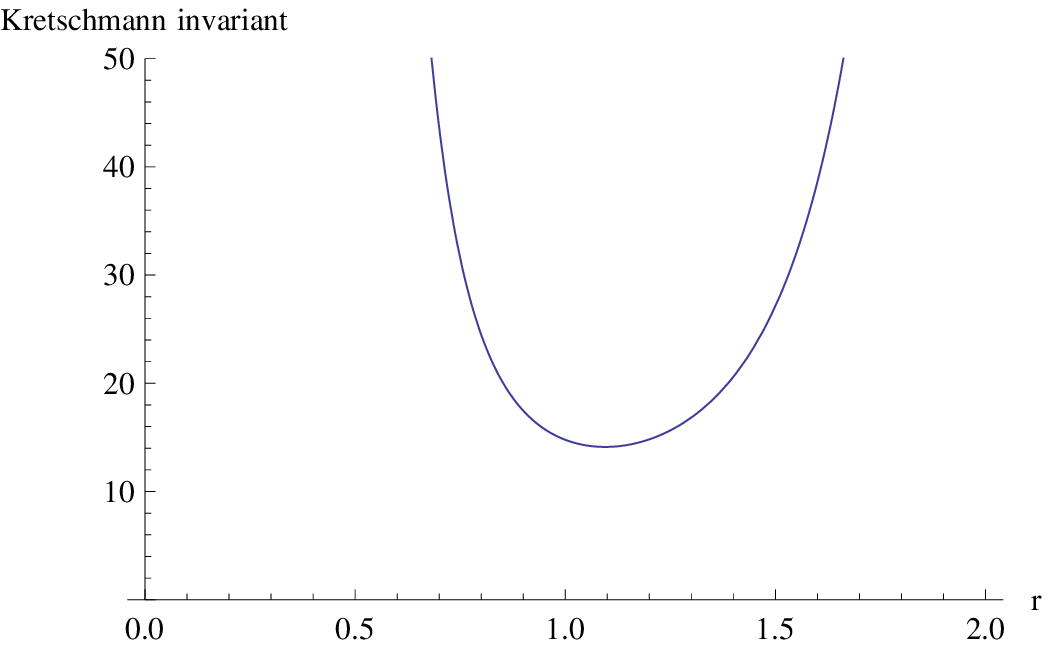}
  \end{center}
  \caption{The Ricci scalar (left) and the Kretschmann invariant (right) 
  for the metric (\ref{matrixgrav1}). Here we set $\theta =1$.} 
  \label{fig:rs_diagonal_metric}
\end{figure}
All of them diverge both at $r=0$ and $r=\infty$, where $\det G(r) =0$.
From Figure \ref{fig:rs_diagonal_metric} we see that 
the ``interior" region is actually $1.0 < r/\sqrt{\theta} < 1.5$, 
and in this region
the spacetime seems to be a ``warped" Minkowski space 
in the sense that the scalar curvature behaves 
as if it is almost constant and the metric has an overall scaling factor 
$(4r^2/\theta) e^{-r^2/\theta}$. 
These analyses indicate that this spacetime is seen as a small bubble of ordinary space 
surrounded by the empty space (nothing state).
The size of this bubble is approximately $\sqrt{\theta}$.

\subsection{Generalizations}

The above solution can be immediately generalized in two ways.

First, we can change the choice of the two indices from $i=0,1$ to any other pair, 
because it is not important to construct a solution.
In particular,  this generalization is related to 
the solution-generating technique \cite{Harvey:2000jb}.
Note that if $E_\mu^a$ is a solution then 
$SE_\mu^a S^\dagger$ is also a solution\footnote{
Note that the unitary transformation $E_\mu^a \to U E_\mu^a U^{-1}$ is a 
symmetry of the cosmological action.}.
Here $S$ is a shift operator defined by
\be
S=\sum_{i=0}^{\infty} \ket{i+1}\bra{i}
\ee
and satisfies $S^{\dagger} S=1$ but $SS^\dagger=1-\phi_0$.
Thus, the metric
\begin{align}
G_{\mu\nu}&= \eta_{\mu\nu} S \left(\ket{0}\bra{0}+\ket{1}\bra{1}\right) S^\dagger 
\nonumber\\
&= \eta_{\mu\nu} S \left(\ket{1}\bra{1}+\ket{2}\bra{2}\right) S^\dagger 
\nonumber\\
&=\eta_{\mu\nu} \left(\phi_1 + \phi_2  \right)
\end{align}
is also a solution.
The choice of the two indices does not affect the size of the ``bubble" 
on the noncommutative space because $\Tr\left(\phi_1 + \phi_2  \right)=2$ is 
the same as above.

Next generalization is to enlarge the size of gamma matrices.
To do this, choose the index $i=0,1,\cdots, q$, where $q=2^{[d/2]}-1$ and 
define $SO(d)$ gamma matrices $\gamma^0, \cdots, \gamma^d$ in the harmonic oscillator 
space as above.
Then by selecting three of them, say, 
\be
E_0^0=\gamma^0, \quad E_1^1=\gamma^1, \quad E_2^2=\gamma^2 ,
\label{local Min vielbein}
\ee
they also solve the equation of motion.
Because now the size of gamma matrices is $2^{[d/2]}$, the corresponding metric 
is proportional to the rank $2^{[d/2]}$ projection operators as
\begin{align}
G_{\mu\nu}&= \eta_{\mu\nu} 
\left(\ket{0}\bra{0}+\cdots +\ket{q}\bra{q}\right) \nonumber\\
&=\eta_{\mu\nu} \left(\phi_0 + \cdots +\phi_{q}  \right).
\label{local Min metric}
\end{align}
The volume of the support which will contribute to the effective cosmological constant term
is $\Tr \left(\phi_0 + \cdots +\phi_{q}  \right) = q+1$ times larger than previous 
solutions, as expected.
In particular, by taking a large matrix-size limit $q\to \infty$,
we find that (\ref{local Min metric}) 
actually reduces to the Minkowski metric because of 
the completeness relation $\sum_{i=0}^{\infty} \phi_i =1$ (\ref{completeness}).
The derived second-order cosmological constant term in this limit
\be
S=-2\lambda \int\! dtd^2 x
\ee
is in fact divergent.
It is surprising that the Minkowski spacetime can emerge only from 
the cosmological constant term.
And it is rather confusing that the Minkowski metric carries the divergent 
cosmological constant,
because that spacetime is a classical vacuum for the vanishing cosmological constant
in the ordinary sense.
The point is that we see the spacetime from the nothing $G_{\mu\nu}=0$ 
as the ground state, where the Minkowski space has infinite volume, while in the 
ordinary Einstein equation it is implicitly assumed that
the Minkowski space is a ground state.
Therefore it is not a contradiction.
In summary, we found a sequence of solutions that interpolates 
$G_{\mu\nu}=0$ ($q=0$) and the Minkowski space ($q\to \infty$).

Another interesting application is to choose the index now 
starting from $1$, i.e., $i=1,\cdots, q$ with $q=2^{[d/2]}$
and to take $q\to\infty$ limit.
Then $E_0^0=\gamma^0$, $E_1^1=\gamma^1$, $E_2^2=\gamma^2$ 
in this basis define a solution as above
but now the metric becomes 
\begin{align}
G_{\mu\nu}&= \eta_{\mu\nu} 
\left(\ket{1}\bra{1}+\cdots +\ket{q}\bra{q}\right) \nonumber\\
&=\eta_{\mu\nu} \left(\phi_1 + \cdots +\phi_{q}  \right)\nonumber\\
&\xrightarrow{q\to\infty}  \eta_{\mu\nu} \left(1-\phi_0 \right). 
\label{sing Min metric}
\end{align}
Thus, the metric approaches to $G_{\mu\nu}=(1-2e^{-r^2/\theta})\eta_{\mu\nu}$. 
More generally, by choosing the index $i=k,\cdots, \infty$ for some $k$, 
we have a class of solutions 
\begin{align}
G_{\mu\nu}
&\xrightarrow{q\to\infty} \eta_{\mu\nu} \left(1-\phi_0-\cdots-\phi_k \right).
\label{sing Min metric k}
\end{align}
As opposed to the solutions above, the metric (\ref{sing Min metric}) has
a support for all over the spacetime except for that of $\phi_0$.
Then as seen from the ``noncommutative" determinant,
this spacetime is seen as a ``hole" of minimal size $\sim \sqrt{\theta}$ 
in the Minkowski space.
Similarly, the metric (\ref{sing Min metric k}) has a hole of radius $k\sqrt{\theta}$ 
in the 
Minkowski spacetime.

It is also seen from the analysis of ``commutative" quantities.
Indeed, a ``hole" for the metric (\ref{sing Min metric}) is roughly seen by
its ``step function" profile (see Fig.\ref{fig:1-phi0}) jumping at 
$r=\sqrt{(\ln 2) \theta}\sim 0.833\sqrt{\theta}$, 
which is the zero point of $\det G$.
\begin{figure}[t]
  \begin{center}
    \includegraphics[scale=0.65]{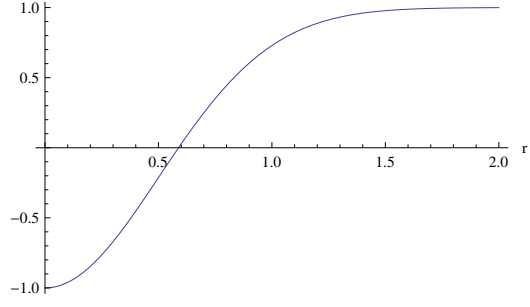}
  \end{center}
  \caption{The profile of $1-2e^{-r^2/\theta}$. Here we set $\theta=1$.} 
  \label{fig:1-phi0}
\end{figure}
Checking now the invariant scalars of this spacetime (\ref{sing Min metric}),
we find the concrete forms of them as
\ba
&& R = -\frac{8e^{r^2/\theta}}{(e^{r^2/\theta}-2)^3 \theta^2}
\left\{\left(1-2e^{r^2/\theta}\right)r^2
+2\left(-2 + e^{r^2/\theta}\right)\theta \right\} \\
&&R_{\mu\nu\rho\sigma}R^{\mu\nu\rho\sigma}
= \frac{32e^{2r^2/\theta}}{(e^{r^2/\theta}-2)^6\theta^4} \nn
&&\times\left\{\left(2+4e^{2r^2/\theta}\right)r^4
-2\left(2-7e^{r^2/\theta}+3e^{2r^2/\theta}\right)r^2\theta
+3\left(-2 + e^{r^2/\theta}\right)\theta^2 \right\} 
\ea
Both of them are finite at the origin and 
diverge at $r \sim 0.833 \sqrt{\theta}$. 
\begin{figure}
\begin{center}
\includegraphics[scale=0.65]{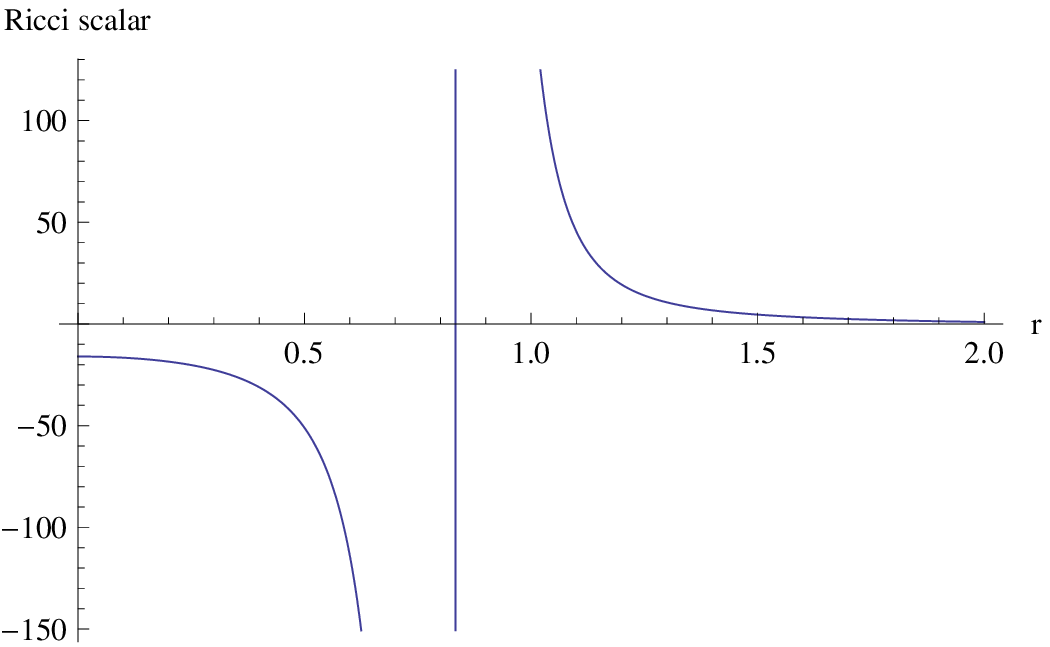}
\quad
\includegraphics[scale=0.65]{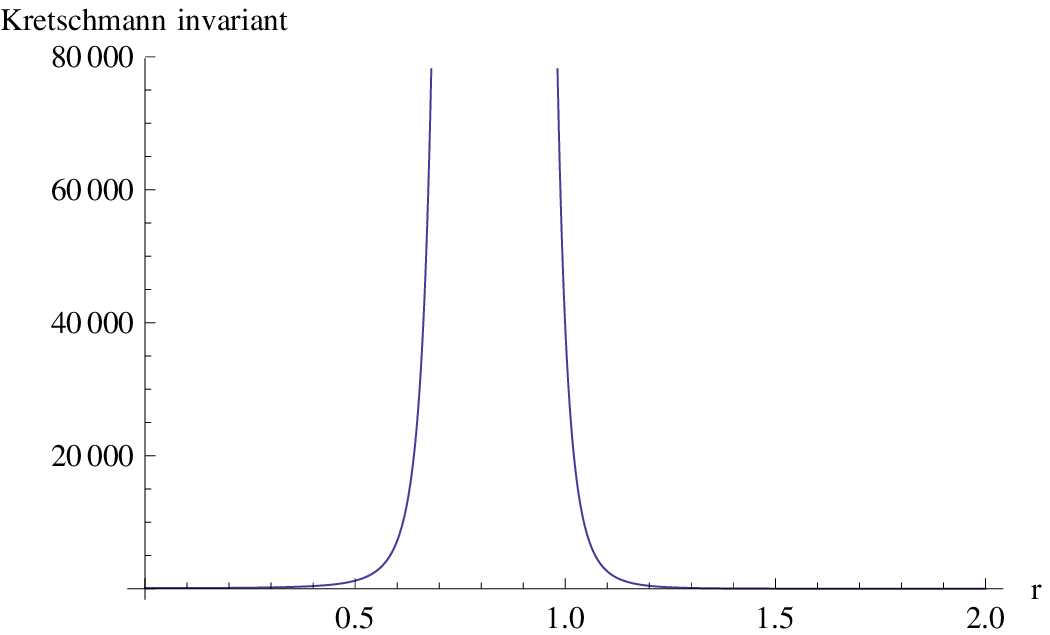}
\caption{The Ricci scalar (left) and the Kretschmann invariant (right)
of the spacetime (\ref{sing Min metric}) for $\theta =1$.}
\label{sing Min metric_fig}
\end{center}
\end{figure}
As shown in Fig.\ref{sing Min metric_fig},  
the spacetime are divided into two regions by a wall 
where the curvature diverges. 
Therefore, 
two bubbles of the universes of different curvatures seem to be glued 
at the curvature wall. 
The outer region has an almost zero scalar curvature so that it 
is expected to be the Minkowski spacetime outside the ``hole", 
in order to be consistent with the ``noncommutative" quantities.
On the other hand, the interior region has a negative scalar curvature.
This naively indicates the AdS spacetime, which is not expected from the 
noncommutative viewpoint.
However, we would not like to be serious about the precise value of the scalar curvature
in this ``commutative" evaluation.

A remarkable feature of this class of solutions
(\ref{local Min metric}), (\ref{sing Min metric}) and (\ref{sing Min metric k})
is that they do not shrink to a point in the commutative limit. 
In the limit, all the projection operators $\phi_i$ reduce to sharp, 
delta-function type distributions, 
and all the degenerate points with $\det_\star G=0$ and $\det G=0$ 
will concentrate to the origin.
Then the metrics (\ref{sing Min metric}) and (\ref{sing Min metric k}) approach 
to the Minkowski spacetime except for the origin, which is the
point-like curvature singularity\footnote{
On the other hand, the metric (\ref{local Min metric}) approaches to
spacetime with a single point around the nothing.}.
Indeed, it is easily shown that $R \to 0$ for $r\ne 0$ 
and $R \sim 16/{\theta} \to \infty$ for $r=0$.
It is again interesting to see this as the resolution of singularities.
When the spacetime is highly curved at a point 
in the second-order formulation of gravity,
one would expect strong effects of quantum gravity to appear at this point.
In our case, it is simply the degenerate point of the metric, where 
the second-order formulation becomes meaningless but they are 
still well defined in the first-order noncommutative formulation.
This scenario is analogous to the stringy resolution of singularity 
in the instanton moduli space, where the small instanton singularity 
is not an end of the moduli space actually and is connected to other 
branches of the vacua.
Moreover, each singularity carries a kind of an index ($k$ in (\ref{sing Min metric k})),
which has a definite meaning in the noncommutative space as the size of the singularity.
This reminds us to the black hole microstates.

%-----------------%
\section{Conclusions and Discussions}

In this paper, we investigated the three-dimensional gravitational theory 
on the noncommutative space. 
We considered the setting where the action has the cosmological 
constant term only.
Although the action has no kinetic term, 
we found infinitely many nontrivial classical solutions 
owing to the noncommutativity. 
In order to construct the solutions, we applied the recipe 
developed in \cite{Gopakumar:2000zd}, i.e., the usage of 
the connection between the star product and the operator formulation. 

To understand the solutions, we proposed a new point of view for the 
cosmological constant term,
that is, the action we gave here is already a full theory
without introducing the scalar curvature term.
When we adopt this idea, the metric, the Ricci scalar and other 
physical quantities can be constructed after the vielbein 
is obtained by solving the equation of motion (\ref{eom}). 
In other words, we switch the second-order formalism effectively.
In this case, the vielbein which solves (\ref{eom}) can be 
regarded as a ``meta" spacetime or a seed vielbein that 
can work as a source for the commutative Ricci scalar $R$ or 
the noncommutative one $R_\star$.  
We would like to emphasize that this point of view has never 
appeared. 
One of the reasons for that is that on commutative spaces, 
a cosmological constant term itself can not give a nontrivial
solution, but it needs a kinetic term. 

Let us now summarize the solutions that are classified into two classes. 
The solutions of the first class are constructed by using the projection 
operators. 
We constructed general solutions of this class.
All of them satisfy $\det_\star G =0$ but $\det G \ne 0$, so we  
calculated commutative scalar curvatures produced by the metrics 
based on the solutions of the vielbein. 
We found that the spacetimes divided into several regions 
by the walls of the curvature singularities where $\det G$ becomes zero. 
In that sense, they have structures of the bubbles of spacetimes with 
various cosmological constants. 
Another feature of this class is that they indicate dimensional reduction, 
that is,  
there are some solutions which are effectively 
one or two dimensional because of the degeneracy $\det_\star G =0$.
In the context of quantum gravity, 
the possibility of dimensional reduction has been 
intensively discussed \cite{Horowitz:1990qb, Ambjorn:2005db} 
or in the other gravitational theory, a similar issue has been 
reported \cite{Horava:2009if}. 
It would be interesting to investigate the relation of our theory to them. 

The solutions of the second class are constructed by applying 
the Clifford algebra and the gamma matrices.
They are noncommutative solitons
interpolating $G_{\mu\nu}=0$ and $G_{\mu\nu}=\eta_{\mu\nu}$.
They satisfy $\det_\star G \ne 0$ and $\det G \ne 0$, so both
noncommutative and commutative quantities can be derived from the vielbein.
This analysis indicates that the solutions are regarded as either
a bubble of ordinary spacetime around the nothing $G_{\mu\nu}=0$, or
a hole (bubble of nothing) in the Minkowski space, 
where their regions with different scalar curvatures are 
partitioned by the wall of the curvature singularity. 
Interestingly, the Minkowski metric is included in this class of solutions,
in which the curvature singularities are absent 
in the large size limit of the gamma matrices.
We also argued the possible mechanism for the 
resolution of point-like curvature singularities in the commutative limit.

Thus we found a lot of nontrivial solutions 
which can be expected to have effects of quantum gravity, 
but there are many open questions to be investigated. 
We would like to note again that 
they depend on the two possible interpretations of the model
discussed in \S 2.

The first possibility is to regard the action we have used in this paper 
as a part of a full theory. 
In other words, we need to add a (noncommutative) spin connection to our theory.
Along this interpretation, the solutions in this paper
would not exact solutions in the full theory.
However, they should be valid in a certain limit where 
the spin connection term is negligible compared with 
the cosmological constant term. 
It is interesting if the existence of our solutions 
would restrict possible noncommutative extensions of the first-order formulation of gravity.
Note that for noncommutative scalar field theories, the solutions 
obtained in the large noncommutativity limit can be extended to the so-called 
exact noncommutative solitons in the full theory with kinetic term 
by adding noncommutative gauge fields.
The spin connection would play a similar role as gauge fields.
It would also be useful to focus on the symmetry of our solutions for that purpose.
We refer that the $E_\mu^a=0$ solution preserves the full (twisted) diffeomorphism, 
while the Minkowski metric preserves the twisted Poincar\'e symmetry.
What is the corresponding twisted symmetry in our case?
Because of the static, rotational symmetric ansatz, a naive guess is 
the twisted version of ${\mathbb R}\rtimes U(1) \times SO(2)$.

Looking at our model from the observational point of view is 
very interesting as well. 
Concerning it, we note that 
there is an argument that $G_{\mu\nu}=0$ is an origin of 
the dark matter \cite{Banados:2007qz}.
Here the $E^a_\mu=0$ does not constrain the spin connection and thus in the equation of 
motion for the fluctuation there is an extra integration constant, 
which behaves as the dark matter.
If such a possibility would be applicable to our solutions as well, 
we might be able to see noncommutative effect by cosmological observations. 
In that sense, we need more ``realistic" solutions, e.g., 
a four-dimensional and time-dependent solution.  
The application of our model to black holes on noncommutative spaces 
is also an interesting direction. 
In the commutative limit $\theta\to 0$ of (\ref{sing Min metric}), 
there appears a sharp, delta-function 
like singularity at the origin which behaves as a point-like source.
There are black hole solutions on the noncommutative space with that kind of 
source term (smeared by the noncommutativity) \cite{Nicolini:2009gw, Nicolini:2005vd, 
Ansoldi:2006vg, Spallucci:2008ez, Myung:2008kp}.
It is interesting to investigate the relation to our solutions.
In the weakly noncommutative case, the $\theta$-expansion works so that 
we can approximately use ordinary Einstein-Hilbert action.
Note that for any finite $k$ the metric (\ref{sing Min metric k}) also represents 
a point-like source but now with $k$ internal degrees of freedom.
There might be a relation to black hole microstates. 

On the other hand, when we regard our theory as a full theory, 
the most important issue is to show the validity of this approach,
in other words, to show the relation to the second-order 
formalism without spin connection. 
Concerning this, we remind that there is already a similar situation 
in string field theory and in the context of quantum gravity
\cite{Horowitz:1990qb}:
There exists the solution which satisfies $\Phi_0*\Phi_0=0$ 
of the pre-geometrical action $S\sim \int \Phi *\Phi *\Phi$ defines 
a BRST charge as $\Phi_0$ and the fluctuation theory becomes Witten's SFT.
This seem to be a very interesting scenario, 
if there is an analogous mechanism for our solutions 
to emerge gravity starting from the cosmological constant only. 

This is the first paper that suggest the emergence of ``meta" spacetimes 
only from a cosmological constant and noncommutativity. 
Besides the ordinary expectation that the noncommutativity becomes important at the
Planck scale, our model may suggest a more radical scenario that 
the noncommutativity would also be crucial for spacetimes even at a large scale. 
In this respect, this scenario gives also a new direction about the 
cosmological constant problem, that is, the cosmological constant 
is necessary for spacetimes to emerge.   
Both fundamental and phenomenological questions on this model 
have to be investigated further.

%%%%%%%%%%
\section*{Acknowledgements}
The authors would like to thank 
Y.~Sasai, 
J.~Soda, 
H.~Ujino, 
H.~Usui 
and S.~Watamura
for helpful discussions. 
This work of S. K. is supported 
by JSPS Grand-in-Aid 
for Young Scientists (B) 21740198. 

%%%%%%%%%%

\appendix

\section{The explicit forms of the Ricci scalar and 
the Kretschmann invariant for the metric (\ref{diagonal_metric})}
\label{Ricci_and_Kretschmann}

We give the explicit forms of the Ricci scalar and 
the Kretschmann invariant for the metric (\ref{diagonal_metric}). 
They are given by
\begin{multline}
R=2 e^{\frac{r^2}{\theta }} 
 \Big\{
 8 x^{12}+x^{10} \left(40
   y^2-68 \theta \right)+8 x^8 \left(10 y^4-37 y^2 \theta +24 \theta
   ^2\right) \\
 \hspace{1cm}\shoveleft+4 x^6 \left(20 y^6-126 y^4 \theta +172 y^2 \theta ^2-65
   \theta ^3\right)\\
 \hspace{1cm}\shoveleft+2 x^4 \left(20 y^8-208 y^6 \theta +456 y^4 \theta
   ^2-359 y^2 \theta ^3+89 \theta ^4\right)\\
 \hspace{1cm}\shoveleft+x^2 \left(8 y^{10}-164 y^8
   \theta +528 y^6 \theta ^2-656 y^4 \theta ^3+330 y^2 \theta ^4-65
   \theta ^5\right)\\
 \hspace{1cm}\shoveleft+\theta  \left(-24 y^{10}+112 y^8 \theta -198 y^6
   \theta ^2+152 y^4 \theta ^3-61 y^2 \theta ^4+10 \theta
   ^5\right)\Big\} \\ 
 /\left\{\theta  \left(-2 r^2+\theta \right)^2 
 \left(2r^4 -4r^2 \theta+ \theta
   ^2\right)^2 \right\}, 
\end{multline}
\begin{multline}
R_{\mu\nu\rho\sigma}R^{\mu\nu\rho\sigma} \\
\hspace{0.5cm}\shoveleft = 4 e^{\frac{2r^2}{\theta }} 
\Big\{
64 x^{24}+64
   x^{22} \left(10 y^2-13 \theta \right)+16 x^{20} \left(180 y^4-476
   y^2 \theta +305 \theta ^2\right)\\
   \hspace{1cm}\shoveleft+64 x^{18} \left(120 y^6-486 y^4
   \theta +639 y^2 \theta ^2-260 \theta ^3\right)\\
   \hspace{1cm}\shoveleft+32 x^{16} \left(420
   y^8-2328 y^6 \theta +4744 y^4 \theta ^2-3989 y^2 \theta ^3+1148
   \theta ^4\right)\\
   \hspace{1cm}\shoveleft+16 x^{14} \big(1008 y^{10}-7224 y^8 \theta +20488
   y^6 \theta ^2-26914 y^4 \theta ^3+16049 y^2 \theta ^4-3487 \theta
   ^5\big)\\
   \hspace{1cm}\shoveleft+8 x^{12} \left(1680 y^{12}-15120 y^{10} \theta +56812 y^8
   \theta ^2-104748 y^6 \theta ^3+97688 y^4 \theta ^4-43900 y^2 \theta
   ^5+7587 \theta ^6\right)\\
   \hspace{1cm}\shoveleft+8 x^{10} \big(960 y^{14}-10752 y^{12}
   \theta +52640 y^{10} \theta ^2-129556 y^8 \theta ^3+169222 y^6
   \theta ^4 \\
   \shoveright{-118388 y^4 \theta ^5
   +42087 y^2 \theta ^6-6071 \theta
   ^7\big)}\\
   \hspace{1cm}\shoveleft+4 x^8 \big(720 y^{16}-10176 y^{14} \theta +65744 y^{12}
   \theta ^2-211400 y^{10} \theta ^3\\
   \shoveright{+365560 y^8 \theta ^4-355060 y^6
   \theta ^5+194909 y^4 \theta ^6-57396 y^2 \theta ^7+7245 \theta
   ^8\big)}\\
   \hspace{1cm}\shoveleft+4 x^6 \big(160 y^{18}-3024 y^{16} \theta +27232 y^{14}
   \theta ^2-114072 y^{12} \theta ^3+252764 y^{10} \theta ^4\\
   \shoveright{-320300 y^8
   \theta ^5+241376 y^6 \theta ^6-108758 y^4 \theta ^7+27825 y^2 \theta
   ^8-3176 \theta ^9\big)}\\
   \hspace{1cm}\shoveleft+x^4 \big(64 y^{20}-1984 y^{18} \theta
   +28688 y^{16} \theta ^2-157984 y^{14} \theta ^3+438784 y^{12} \theta
   ^4\\
   \hspace{1cm}\shoveleft-696832 y^{10} \theta ^5
   +675216 y^8 \theta ^6-413272 y^6 \theta
   ^7+160576 y^4 \theta ^8-36972 y^2 \theta ^9+3897 \theta
   ^{10}\big)\\
   \hspace{1cm}\shoveleft-2 x^2 \theta  \big(64 y^{20}-2208 y^{18} \theta +16112
   y^{16} \theta ^2-54952 y^{14} \theta ^3+106080 y^{12} \theta
   ^4\\
   \shoveright{-126580 y^{10} \theta ^5+98472 y^8 \theta ^6-51586 y^6 \theta
   ^7+17960 y^4 \theta ^8-3815 y^2 \theta ^9+373 \theta
   ^{10}\big)}\\
   \hspace{1cm}\shoveleft+\theta ^2 \big(320 y^{20}-3008 y^{18} \theta +12256
   y^{16} \theta ^2-27984 y^{14} \theta ^3+39812 y^{12} \theta ^4\\
   \shoveright{-37688
   y^{10} \theta ^5+24916 y^8 \theta ^6-11652 y^6 \theta ^7+3741 y^4 \theta ^8
   -738 y^2 \theta ^9+66 \theta ^{10}\big)
\Big\}}\\
   /\left\{\theta ^2 \left(-2 r^2+\theta \right)^4
   \left(2r^4 -4r^2 \theta + \theta ^2\right)^4 \right\},
\end{multline}
respectively.

%%%%%%%%%%%%%%%%%%%

\bibliographystyle{JHEP}
\bibliography{refsAK1}

\end{document}